# Monitoring physical function in patients with knee osteoarthritis using data from wearable activity monitors

Vibhu Agarwal[¶1*], Matthew Smuck[¶2], Nigam H. Shah[¶3]

[1]Biomedical Informatics Training Program, Stanford University, Stanford, USA

[2]Department of Orthopaedic Surgery, Stanford University, Stanford, CA, USA;

[3]Stanford Center for Biomedical Informatics Research, Stanford University, Stanford, USA

*Corresponding author

Email: vibhua@stanford.edu

[¶]Authors contributed equally to this work

# Abstract

## Background

Currently used clinical assessments for physical function do not objectively quantify daily activities in routine living. Wearable activity monitors enable objective measurement of routine daily activities, but do not map to clinically measured physical performance measures.

## Methods

We represent physical function as a daily *activity profile* derived from minute-level activity data obtained via a wearable activity monitor. We constructed the daily activity profiles--representing average time spent in a set of activity classes over consecutive days--using the Osteoarthritis Initiative (OAI) data. Using the daily activity profile as input, we trained statistical models that classify subjects into quartiles of objective measurements of physical function as measured via the 400m walk test, the 20m walk test and 5 times sit stand test. We evaluated model performance on held out data from the same calendar year as that used to train the models as well as using activity data two years into the future.

## Results

The daily *activity profile* derived from minute-level activity data obtained via a wearable activity monitor can accurately predict physical performance as measured via clinical assessments. Using held out data, the AUC obtained in classifying performance values in the 1st quartile was 0.79, 0.78 and 0.72, for the 400m walk, the 20m walk and 5 times sit stand tests. For classifying performance values in the 4$^{th}$ quartile, the AUC obtained was 0.77, 0.66 and 0.73 respectively. Evaluated on data from two years into the future, for the 20m pace test and the 5 times sit stand tests, the highest AUC obtained was 0.77 and 0.68 for the 1$^{st}$ quartile and 0.75 and 0.70 for the 4$^{th}$ quartile respectively.

## Conclusions

We can construct activity profiles that represent actual physical function as demonstrated by the relationship between the activity profiles and the clinically measured physical performance measures. Such objective and continuous measurement of physical performance via the activity profile can enable remote functional monitoring of patients.

## 1 Introduction

Physical function is among the most common indicators of physiological well-being. For example, aging or disease-induced loss of muscle and neural control leads to a reduction in maximal strength. Similarly, reduced cardiac output results in lower aerobic performance. Initial interest in functional limitations that impact daily living was motivated primarily by rheumatology[1–4] and cardiovascular research[5–7]. More recently, physical function status is an outcome of interest within most medical specialties[8–16] and increasingly, is being regarded as the "sixth vital sign"[17].

Endeavors for arresting functional decline must start with an evaluation of plausible interventions against the baseline functional status. For example, the problem of maximizing functional improvements in patients with advanced Osteoarthritis via a surgical or a non-surgical intervention requires knowing both the baseline function as well as subsequent improvement. Therefore, monitoring of physical function through valid metrics is necessary for optimal management of patients. Such monitoring can also be useful in treating conditions where physical function impairment is not the main complaint[18–20].

Physical function is characterized by the ensemble of daily activities an individual is capable of engaging in. The basic set of activities comprising bathing, dressing, toileting, transfer, continence, and feeding are collectively referred to as Activities of Daily Living (ADL) [21,22].

When moderate to severe disability is present, ADLs succeed in capturing the physical function changes that accompany disease progression. For example, disability indexes based on ADLs can differentiate healthy ageing patients, patients with mild cognitive impairment and patients with dementia in a diagnostic setting[23]. However, ADL scores may have a response bias from self-reporting and have low sensitivity to changes in high functioning older adults[24].

By contrast, physical performance measures (such as walking speed, sit to stand speed and grip strength) capture the variation across a wider range of physical function, including changes in the initial stages of decline[25–27]. Subjects taking physical performance tests, attempt predefined tasks that are linked to functional abilities (such as mobility, lower limb strength and overall muscle strength) under the supervision of a test administrator. The chief drawback of performance measures is that they require substantial time and effort of the subjects and the test administrator along with access to specialized skills and facilities. The relationship between physical activity and performance measures of function has been a topic of active research[28–31].

The commonly used self-reported assessments of physical function are vulnerable to biases arising from the perceived desirability of conforming to physical activity norms and from the difficulty in recalling historical activity. By contrast, existing objective measurements of physical function require adherence to specific test protocols and are usually limited to research settings.

In this paper, we focus on a novel method for inferring physical function based on objective measurements of daily physical activity obtained from a wearable device. Unobtrusive and efficient monitoring of daily activity may be carried out through such devices at a population level. Our work enables quantitative monitoring of physical function – the first step towards improved precision in both clinical research and practice.

Wearable activity monitors (WAM), typically equipped with one or more accelerometers, provide a convenient way to objectively measure physical activity [32–38]. However, attempts to study the link between physical activity and physical function via WAMs has thus far been limited[39,40]. van Lummel et al[41] imply that, physical activity and physical performance represent associated but independent domains of physical function, and a change in performance (measured by say, the 6 minute walk test) need not always have a corresponding change in activity levels. They concluded that differentiating physical activity into classes provides a more informative measure of function as compared to single metric based on acceleration.

WAMs produce an aggregate value proportional to the change in velocity over an epoch, which by itself is inadequate for distinguishing activity classes. We hypothesize that patterns in daily activity recorded by a WAM correlate with physical function and therefore, may be used to estimate physical function in units of performance used by the various objective measures of function. We uncover instances of activity classes from daily activity data using an unsupervised approach, and propose the idea of a daily *activity profile*, which represents the mean allocation of time to the different activity classes. Using machine learning techniques, we then classify a subject's daily activity profile, into discrete quartiles of commonly used measures of physical function such as the 400 m walk test.

To the best of our knowledge, a data-driven characterization of daily physical activity as a repertoire of activity classes has not been presented in earlier literature on the subject. We demonstrate the feasibility of discriminating physical function categories with high sensitivity and specificity and discuss potential use cases in medical research and treatment. Figure 1 illustrates our overall workflow.

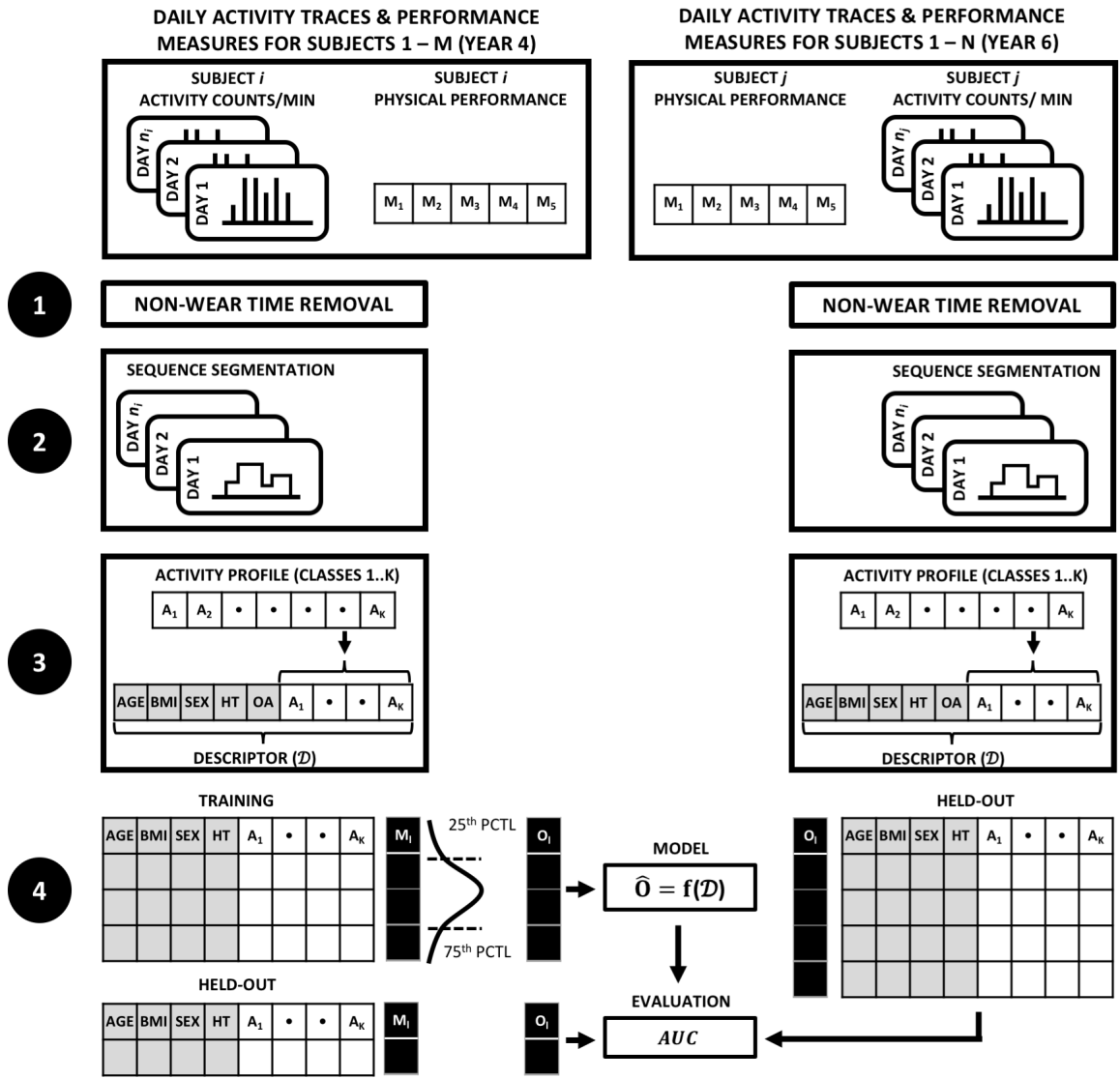

Figure 1. Estimating physical function from daily activity traces. Daily activity was measured as counts per minute was recorded for M=2001 subjects in year 4 along with various objective measurements from which we selected 5 (M1-M5) for our study. A subset of these subjects (N=1406) participated in daily activity and performance measurements again in year 6 1)Non-wear time was excluded from the activity traces 2) The daily activity count sequences were segmented 3) A composite feature descriptor encapsulating the subject's daily activity profile was constructed for each subject. 4) Quantitative response values were converted to ordinal values based on empirical quantiles obtained from the training partition only. A classifier for the feature descriptor was trained on the partition of the feature matrix (80%) and evaluated on the held out partition (20%). The feature matrix obtained through an identical process from year 6 data was used only for evaluation.

# 2 Materials and Methods
## 2.1 Data

We use data from OAI (the Osteoarthritis Initiative, http://www.oai.ucsf.edu), a federally funded initiative for studying a cohort of subjects with clinical Osteoarthritis. OAI provides physical activity data from an invited subset of the subjects who participated in an activity study at the 48 and 72 month follow up visits (year 4 and 6 respectively). The participants in the physical activity studies continously wore an ActiGraph GT1M uniaxial accelerometer (ActiGraph; Pensacola, Florida) for up to seven consecutive days, starting in the morning until retiring at night, except during water activities. The Actigraph GT1M is a compact, hip-worn device that measures dynamic acceleration in the range of 0.05g – 2.0 g, whose validity and reliability is established[42–47]. The participants also maintained a daily log recording time spent in water and cycling activities as the accelerometer may not have been able to capture these activities accurately. Post facto analysis revealed that participants spent little time in water and cycling activities (median 0 minutes/day, interquartile range = 0.0 to 3.4 minutes/day), indicating that little activity was missed by accelerometer monitoring. The key attributes of the 48 month (year and 72-month PA study-data are summarized in Table 1.

Table 1. Key attributes of the physical activity data

|  | Physical activity study | |
|---|---|---|
|  | 48 month | 72 month |
| Number of subjects |  |  |
| Incidence | 1490 | 1080 |
| Progression | 505 | 340 |
| Control | 6 | 6 |

|  | Total | 2001 | 1426 |
|---|---|---|---|
| Gender (% male) | | 44.53 | 45.37 |
| BMI (mean, sd) | | (28.52, 4.87) | (28.22, 4.88) |
| Mean Comorbidity Index | | 0.52 | 0.56 |
| Median days of activity | | 7 | 7 |

The accelerometer data in OAI consists of "activity counts" per minute. An activity counts is a weighted sum of the discretely sampled values of the one-dimensional acceleration. Since zero or low values of activity counts could also arise from non-wear time, we filter out non-wear periods from the data. Continuous runs of zero counts that were more than 90 minutes long (allowing for interruptions of up to 2 consecutive minutes with fewer than 100 counts) were discarded as non-wear periods[48]. A day with a wear time of 10 hours or more was considered as valid. Objective as well as patient reported measures of physical function were recorded for the patients during follow up visits.

## 2.2 Objective Measures of Function

The Osteoarthritis Research Society International (OARSI)[49] recommends a set of tests that examine physical performance in activities that are typically impacted in OA. We selected three performance measures that were used in the OAI study and had equivalents in the OARSI recommended tests: 1) The 400m walk test (400MWT) which is a predictor of mobility disability. Longer completion times on the 400MWT have been shown to be associated with a higher risk of mobility limitation and disability (adjusted hazard ratios 4.43, $p < 0.001$) as well as a higher risk of death (adjusted hazard ratio 3.23, $p < 0.001$) for subjects in the highest quartile[50]. 2) The 20m walk pace (20MPACE), which is the closest available short walk-length evaluation in the OAI dataset. Gait speed, an important functional outcome in knee OA research is generally evaluated using shorter walk-lengths. 3) The 30 second sit to stand test which measures sit-to-stand function in subjects with knee or hip OA[51]. The OAI data provides the number of sit to stands per second measured over 5 repetitions (5CSPACE) as a measure of sit-to-stand function which has good test-retest reliability[52]. In the year 6 follow up, only 20MPACE and 5CSPACE were recorded, whereas all three measures were recorded in year 4.

## 2.3 Relationship with Daily Activity

Physical function is defined based on the repertoire and relative proportion of activity classes that a subject can engage in, within a given environment. We recovered homogenous segments from the daily activity sequences of counts-per-minute and defined activity classes on the basis of similar segments. A subject's daily activity profile was computed as the average minutes allocated to each activity class in that day. Finally, we inferred mappings from the daily activity profile to the objective measurements of function described in the earlier section using supervised learning.

### 2.3.1 Activity Classes and Daily Profile

We used change-point analysis to segment the counts-per-minute sequences. The algorithm searches for segment boundaries such that each segment represents a change in the distribution of the time-ordered counts, with respect to its preceding and following segments. The method described by James and Matteson[53] is non-parametric. We followed the hierarchical divisive method, wherein a change point is obtained via a search over all indexes such that the scaled empirical divergence statistic is maximized. Segments were indexed using the mean and the standard deviation (SD) of the counts-per-minute values (Figure 2). Such a representation has been shown to improve discrimination between activity classes by Pober et al[54].

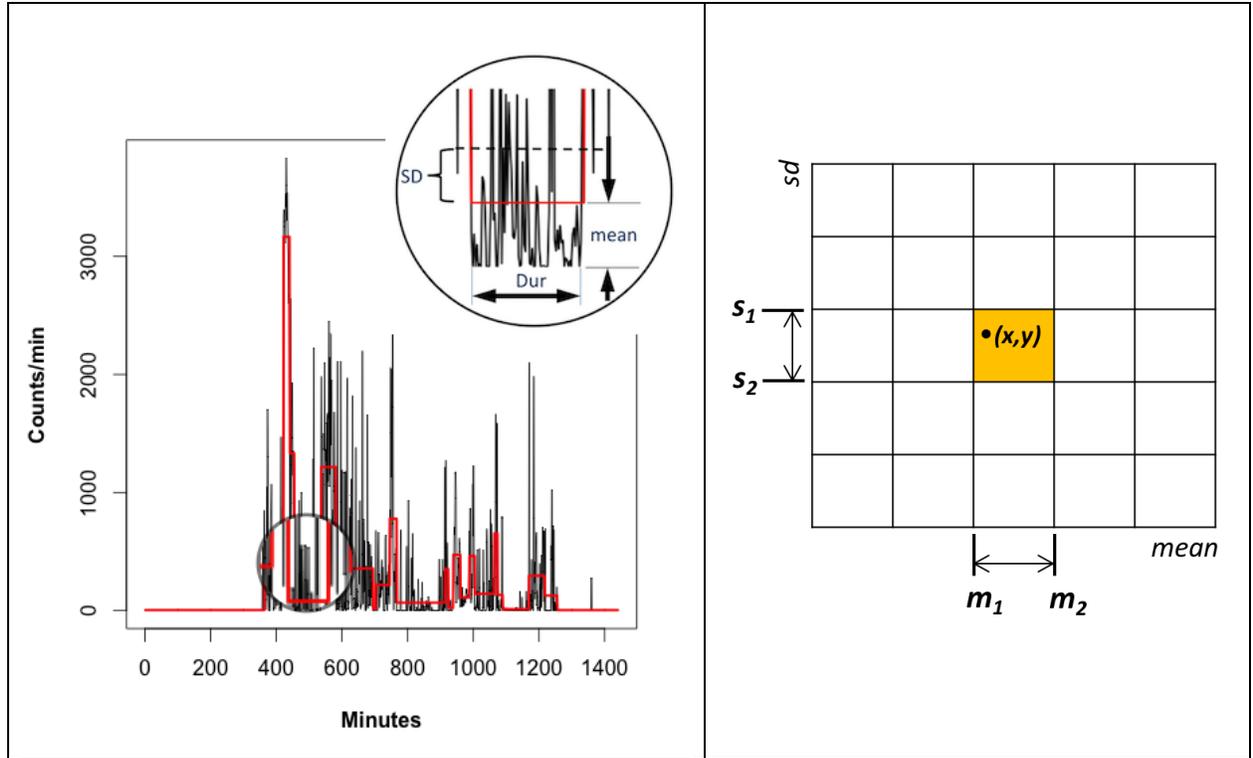

**Figure 2.** A) Segmentation of the counts per minute sequences. Horizontal red lines mark the segments recovered via change point analysis. Each segment is an instance of an activity class whose mean and SD is estimated by the sample mean and SD of the segment. B) The mean and SD space containing all segments is partitioned into bounding regions, each defined by a mean and a SD interval. The shaded region represents activity class ($[m_1, m_2), [s_1, s_2)$).

An activity class is a bounded region in the segment feature space. Our feature space $F$ consists of all $(m, s)$ vectors: $m \in [0, M], s \in [0, S]$, where M and S are the maximum mean and SD over all segments found via the segmentation method. An activity class is defined by a pair of intervals such as $([m_1, m_2), [s_1, s_2))$. A segment with mean $x$ and SD $y$ ($m_1 \leq x < m_2, s_1 \leq y < s_2$) is an instance of the activity class so defined. Based on the activity classes obtained from the partitioning of the segment feature space $F$, we defined a daily activity profile for each subject as the average time allocated to each activity class per day. The activity profile for a subject $i$ is given by

$\mathcal{A}_i = (a_{i1}, a_{i2} \ldots a_{iJ})$ where,

$$a_{il} = \frac{1}{K_i} \sum_{k=1}^{K_i} t_{ilk}$$

$J$: the number of activity classes
$K_i$: the number of days of observations for subject $i$
$t_{ijk}$: the number of minutes spent by subject $i$, in activity class $j$, on day $k$

## 2.4 Supervised Learning

We defined a composite descriptor $\mathcal{D}_i = (BMI_i, Age_i, Sex_i, Height_i, OA_i, \mathcal{A}_i)$ for every subject $i$ in our data where $OA_i$ refers to the categorical status healthy, at-risk or with progressive disease, of subject $i$ at baseline. A regression function $f(\mathcal{D})$ that maps the composite descriptor $\mathcal{D}_i$, to an objective measure of physical performance can be obtained by minimizing the expected squared error loss.

### 2.4.1 Ordinal Response Models

Our activity classes populate a sparse feature space, with most instances occurring in the low mean and low SD regions. A popular technique for dealing with sparse feature spaces is to use non-smooth loss functions[55,56] that allows some of the coefficients to be shrunk to zero. Medical studies commonly group continuous variables into quantiles for ease of interpretation and analysis[50,57,58]. We therefore grouped each of our quantitative response values into ordered categories 1 < 2 < 3 where categories 1 and 3 represent values lying in the lowest and highest quartiles respectively and category 2 represents values spanning the inter-quartile range for the specific response. Continuation ratio logits exploit the sequential nature of the ordering i.e. the observation corresponding to a response value 3 may have earlier had values 1 and 2, and appear well suited for modelling the ordinal categories of the physical performance measurements. An important consequence of the continuation ratio formulation is that the overall multinomial likelihood factors into a product of K-1 binomials, which can be maximized independently (supporting information S1).

The size of the 2D interval in feature space that defines our activity classes is a tuning parameter. Note that the daily activity profile depends on the activity classes, which are defined as intervals in the mean-SD space covering all segments. Small intervals allow instances from adjacent classes to be in close proximity, thereby increasing correlation between the activity profile elements. Interval size selection therefore presents a bias-variance trade-off with small interval sizes leading to high variance in the squared prediction error. We ascertained the optimal size of the bounding region through repeated 5-fold cross validation on our training data (Figure 3). For the optimal bin sizes, The models were refit using the full training data and features based on the optimal bin size. We refer to the resulting L1 penalized ordinal response models with continuation-ratio logits as LPORM models.

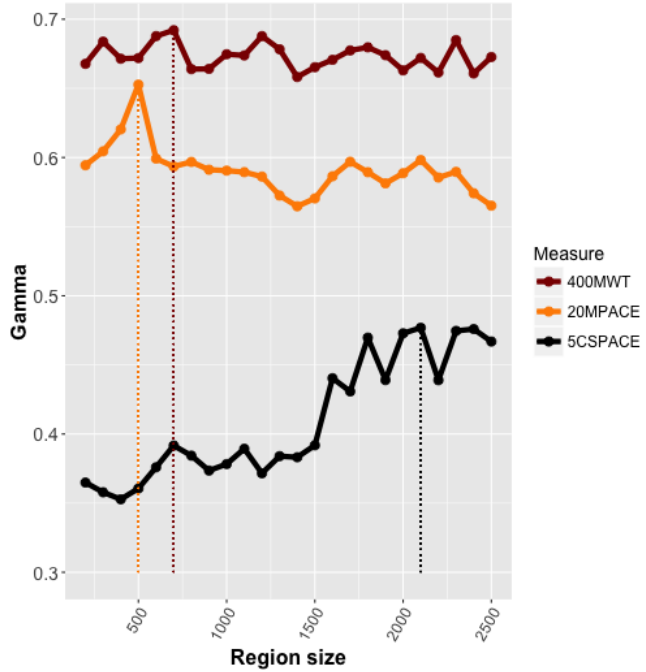

Figure 3. Mean Gamma in 5 fold cross validation for the 20m pace, 5CS and the 400MWT models. The dashed lines indicate the optimal region size

We evaluated the area under the receiver operating curve (AUC) for discriminating category 1 versus others, and category 3 versus others, using the held-out data.

### 2.4.2 Generalized Additive Models

The parameters estimated by LPORMs are biased due to the constraints imposed on the models' degrees of freedom. Consequently, the fitted models are hard to interpret. The fits from the quantitative regressions also suggest that at higher values, linearity in the predictors may not be a justifiable modelling assumption. Generalized Additive Models (GAM) are able to identify and characterize non-linear regression effects through an additive specification of non-parametric functions of the predictors. The unspecified functions of the predictors are smoothers (typically kernels or a cubic splines) that are estimated simultaneously using a backfitting algorithm[59]. A GAM may be specified as follows

$$g(\mu(X)) = \alpha + f_1(X_1) + f_2(X_2) + \cdots + f_p(X_p) \quad \text{where}$$

$\mu(X)$ denotes the conditional mean of the response i.e E[Y|X]

$g$ (.) is the link function

$f_1 \ldots f_p$ are the unspecified smooth functions for each of the p predictors.

The estimated $f_i$ reveal the nature of the relationship of the predictor with the response, an insight that is difficult to obtain through penalized regression. Corresponding to each ordinal response model that we had fit using the L1 penalized continuation-ratio logit, we fit a GAM on our training data using the mgcv package[60], specifying 'ocat' (ordered categorical) for the response family. As before, the AUC values for category 1 and 3 were computed based on the held-out data.

## 3  Results

As described in the methods, we find homogenous segments from the daily activity sequences of counts-per-minute and define *activity classes* on the basis of similar segments. A subject's *activity profile* is the average daily minutes allocated to each activity class. Finally, we inferred mappings from the daily activity profile to the objective measurements of function such as the 400m walk, the 20m walk and the 5 sit to stand tests. The daily activity profile derived from minute-level activity data obtained via a wearable activity monitor can predict physical performance as measured via clinical assessments.

## 3.1 Predicting Ordered Categories

For the daily activity profiles based on the optimal interval sizes, classifier performance for the LPORMs on the held out data is summarized in table 2(A). Inclusion of the activity profile in the LPORMs improves the AUC by 1% – 5% and Gamma by 3%-13% in category classification tasks when evaluated on held out data from year 4, with higher improvement observed for measures based on walking test (400MWT and 20MPACE). Lowest quartile classification in 20MPACE and 5CSPACE improves the least with the inclusion of the activity profile.

Table 2. Gamma and AUC for L1 ordinal regression models evaluated on held out data from Y4. Evaluation results on held out data from Y6 is given within parentheses. Shaded columns represent models with the optimal interval size obtained in cross validation.

| Interval size | 500,500 | | | 700,700 | | | 2100,2100 | | | Without Activity Profile | | |
|---|---|---|---|---|---|---|---|---|---|---|---|---|
| | 400MWT | 20MPACE | 5CSPACE | 400MWT | 20MPACE | 5CSPACE | 400MWT | 20MPACE | 5CSPACE | 400MWT | 20MPACE | 5CSPACE |
| (A) L1 Penalized Ordinal Logistic Regression | | | | | | | | | | | | |
| Gamma | 0.62 | 0.56 | 0.49 | 0.60 | 0.53 | 0.48 | 0.58 | 0.52 | 0.50 | 0.52 | 0.43 | 0.47 |
| AUC 1 | 0.78 | 0.78 (0.77) | 0.72 (0.67) | 0.79 | 0.78 (0.77) | 0.73 (0.67) | 0.78 | 0.77 (0.77) | 0.72 (0.68) | 0.75 | 0.73 (0.75) | 0.70 (0.68) |
| AUC 3 | 0.78 | 0.66 (0.75) | 0.70 (0.68) | 0.77 | 0.67 (0.75) | 0.72 (0.68) | 0.75 | 0.67 (0.75) | 0.73 (0.70) | 0.72 | 0.65 (0.73) | 0.73 (0.71) |
| (B) Generalized Additive Models | | | | | | | | | | | | |
| Gamma | 0.62 | 0.49 | 0.43 | 0.62 | 0.52 | 0.55 | 0.60 | 0.56 | 0.50 | 0.52 | 0.46 | 0.47 |
| AUC 1 | 0.73 | 0.78 (0.49) | 0.69 (0.53) | 0.75 | 0.79 (0.74) | 0.72 (0.67) | 0.78 | 0.78 (0.64) | 0.73 (0.57) | 0.74 | 0.74 (0.76) | 0.72 (0.71) |
| AUC 3 | 0.76 | 0.62 (0.47) | 0.66 (0.53) | 0.77 | 0.65 (0.69) | 0.73 (0.66) | 0.77 | 0.68 (0.61) | 0.72 (0.62) | 0.73 | 0.66 (0.71) | 0.72 (0.71) |

The AUCs for 400MWT model with activity profile based on an interval size of (700,700) imply (at the optimal cut off point) a sensitivity = 0.72, specificity=0.71 for the lowest quartile and a sensitivity = 0.67, specificity = 0.74 for the highest quartile. The optimal cut-off point sensitivity and specificity for classifying the gait speed in the lowest quartile with 20MPACE model, interval size (700,700), are 0.76 and 0.68 respectively. Overall, the concordance is less sensitive to interval size compared to the rank correlation (Gamma).

Table 2(B) shows the model evaluation results for the GAMs with the activity profiles computed on the same interval sizes as the L1 penalized ordinal response models. The performance of the GAMs improved with increase in the interval size and for the largest interval size evaluated, (2100, 2100) the evaluation results on year 4 data for GAMs were comparable with LPORMs.

### 3.1.1 Results on Year 6 data

Evaluation of model performance on year 6 data allows us to assess the performance of our models (trained on year 4 data) on future data, thus emulating non-stationarity in the underlying data. The results in table 3 (in parentheses) show that classification performance for category 1 on held-out data dropped for both 20MPACE and 5CS. However, the LPORM performance in category 3 classification for 20MPACE *improved*. Analysis showed that the marginal distributions of mean daily minutes in the predictor activity classes changed differently from year 4 to 6, for subjects who had a category 3 20MPACE versus others. In particular, the high functional patients on average, spent more time in the activity class ([701,1400), [701, 1400)) in year 6 compared to year 4. The direction of change in the said class for low functional patients was opposite. Overall, the change in the mean daily minutes in the predictor activity classes improves the separation between the 20MPACE high performers and others in year 6 (supporting information S2).

## 3.2 Predictors of Physical Performance

We studied the predictors that were significant (at $p = 0.05$) in the GAMs and were also among the selected predictors in the corresponding LPORM (figure 4). Relationships between the response and predictors in a GAM may be studied via plots of the respective smoothers fitted by the GAMs. These are presented separately for two groups of activity classes, namely activity classes below mean 1400 (figure 5) and activity classes above mean 2800 (figure 6). The activity class ([2100-2800), [2100-2800)) was disregarded on account of disagreement with LPORM selection.

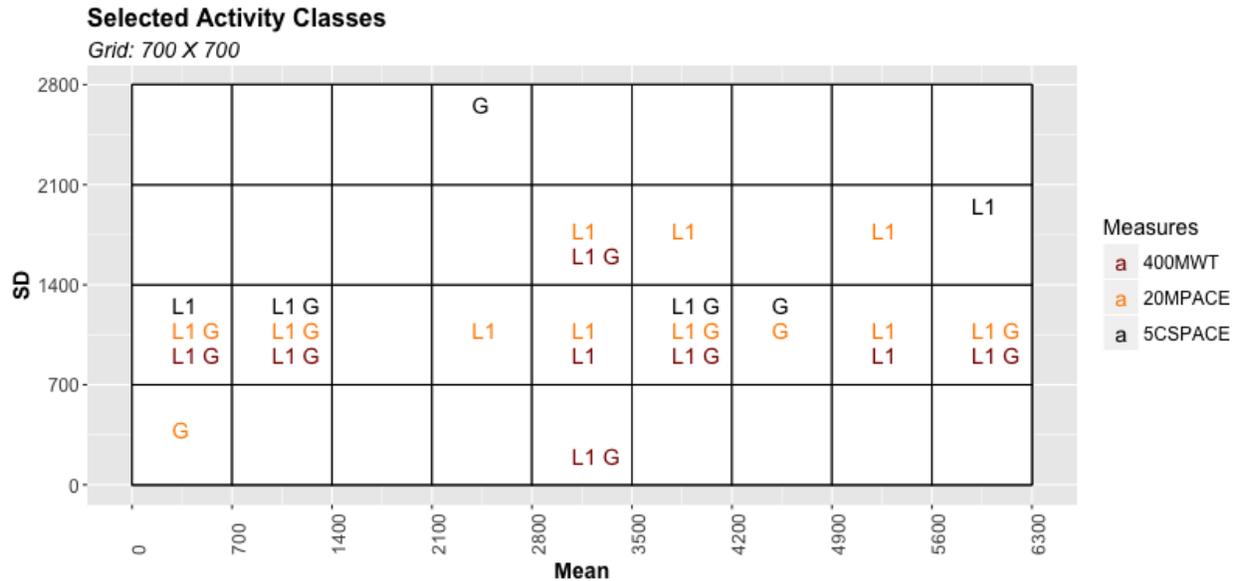

**Figure 4. Each square represents an activity class defined by an interval in the mean, SD space. If the predictor corresponding to the class was selected in a LPORM, it is labelled as L1 and G if the predictor was selected in a GAM. The label colors denote the performance measure for which the predictor was significant.**

Since the 400MWT is a time measurement, it's observed trends are opposite to those of the 20MPACE and the 5CSPACE, which are speed measurements. It is expected that most instances of the activity classes in Fig 5A fall within the cut-points for sedentary or light physical activities (taken as 0 – 2500 activity counts per minute, as defined for populations with mobility limitations by Smuck et al)[61]. The smooth function plots for activity classes ([0,700), [701,1400)) & ([701,1400), [701,1400)) show that up to 25 daily average minutes in these activity classes are associated with improved response in both the 400MWT and the 20MPACE. The 3D scatter plot of all activity class instances (Fig 5B) shows that most activity classes have only a small number of instances.

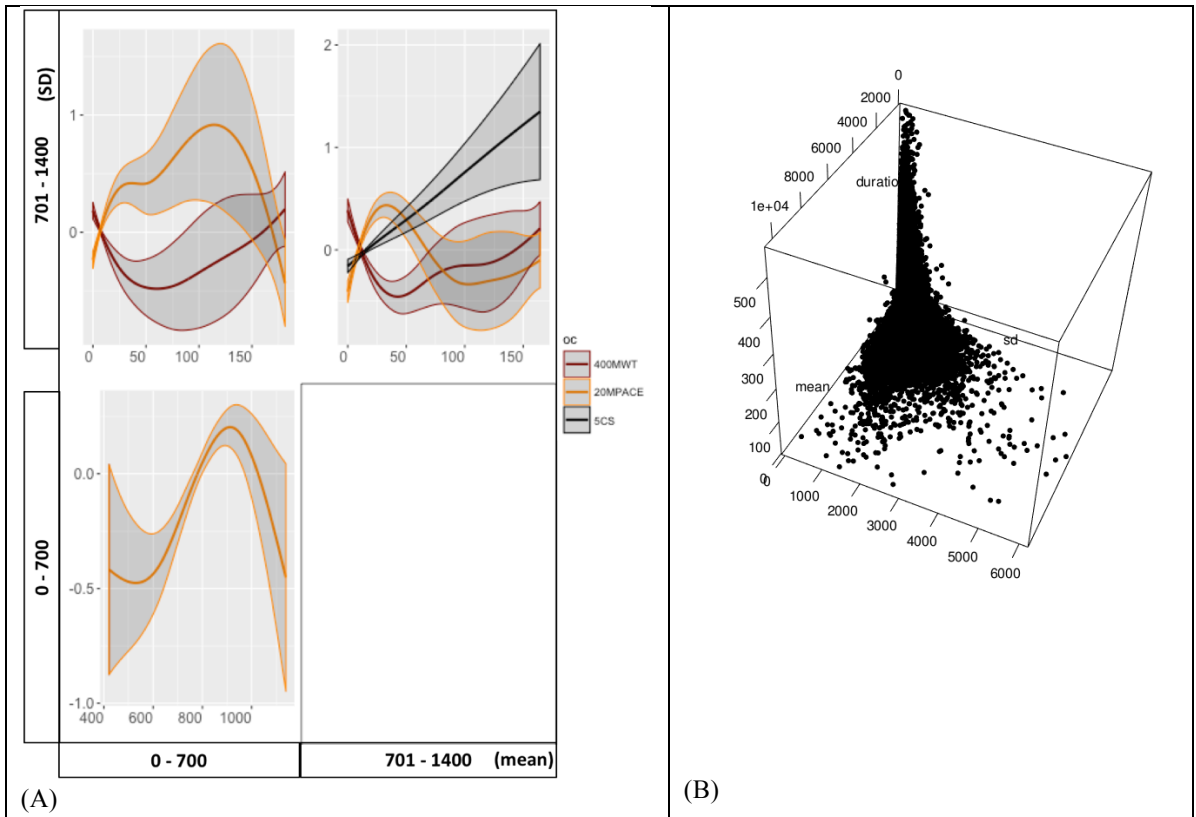

**Figure 5. (A)Smooth functions for low and moderate activities. Each grid square represents an activity class. The smooth functions plotted represent the relationship of the activity class to the (latent) response shown as the latent variable value, as modelled by the ordinal regression on the Y axis and daily average minutes on the X axis. The colors represent different physical performance measures. (B)The 3D scatter plot on the right shows the mean, SD and duration of all activity class instances in our data.**

The number of instances per class decrease quickly as the mean and SD increase resulting in a sparse daily activity profile. Most instances of activity classes that have a higher mean and SD have low duration, resulting in fewer degrees of freedom available for estimating the smooth functions at high values which explains the wider confidence bands for the function estimates at higher values of daily average minutes. Inspection of a sample of instances from the class (701 – 1400, 701 – 1400) revealed that typically, long duration instances are spells of rest punctuated by frequent interruptions. In absence of annotated activity data, it may be reasonable to conjecture that such interruptions involve sit-stand transitions and therefore, the said instances are associated with improved 5CSPACE response. Activity classes with the mean interval (0 – 700) are not associated with 5CSPACE performance. Activity classes ([2800,3500), [0,700)) and ([2800,3500), [1401,2100)) as shown in Fig 6, are associated with 400MWT performance.

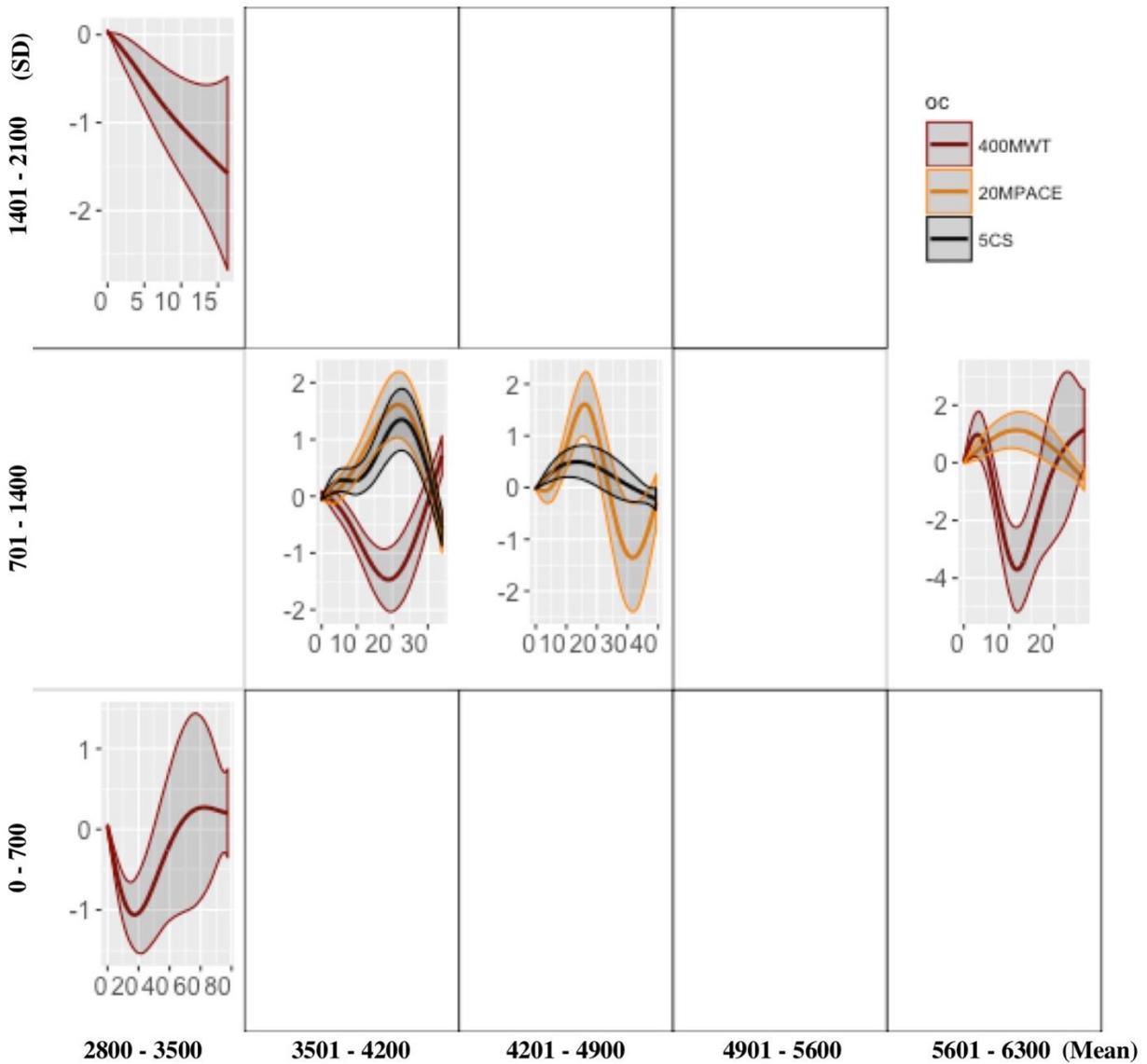

**Figure 6 Smooth functions for moderate to vigorous activities**

Approximations to the distribution of activity counts in any given activity class can be obtained via tail probability bounds. For example, we have from Chebyshev's inequality $P(|X - \mu| > k\sigma) < \frac{1}{k^2}$, where $\mu$ & $\sigma$ may be taken as the mid-interval values of the mean & SD intervals respectively for a given activity class. Substituting the values for the class ([2800,3500, [0,700)) and taking the complementary region probability, we obtain

$P(|X - 3150| < k.350) > \left(1 - \frac{1}{k^2}\right) = 0.69$ for k = 1.8, implying that approximately 70% or more activity counts per minute lie between 2520 and 3780. Thus, most of the activity in the class (2800 – 3500, 0-700) is likely to be in the lower moderate intensity range. Similarly, for the class (2800 – 3500, 1401 - 2100), we note that approximately 70% or more activity counts per minute lie below 6300 which indicates activity in the moderate range along with some vigorous activity. From the corresponding smooth function plots in Fig 6, higher daily average minutes in both classes are associated with improved long walk performance. The association with increased completion times with higher (> 20) daily average minutes in the class ([2800 – 3500),[0-700)) is due to instances comprising of mostly sedentary activity. Infrequent occurrences of such instances also result in wide estimation intervals for the smooth function.

### 3.2.1 Moderate to Vigorous Activity with Malalignment

In the activity class ([3501,4200), [701,1400)) an increase in daily average minutes is initially associated with improved responses in all three performance measures. However, an increase of more than 20 minutes is associated with a decline. Unlike the classes with low mean and SD, instances longer than 20 minutes of the said class do not represent sedentary activity. The decreasing physical function trend with increasing time in moderate to vigorous activity is counter-intuitive. Therefore, we reviewed patient reported outcomes obtained via the Physical Activity Scale for the Elderly (PASE) – an instrument for studying engagement in different kinds of daily activities related to leisure, household and occupational work in the elderly patients[80]. We also reviewed joint exam results reporting varus (bow-legged) and valgus (knock kneed) alignments for the same subjects. These are summarized in table 3 and suggest that the subjects with more than 20 daily average minutes in the activity class ([3501,4200), [701,1400)) have higher prevalence of and severity of knee malalignment, higher time in the activity class (minutes as well as frequency), and fewer sitting hours along with higher walking hours per week.

Table 3. Knee alignment and PASE results of subjects with at least one instance of the activity class (3501-4200, 701-1400)

| Daily average minutes | >= 20 | < 20 |
|---|---|---|
| Number of subjects | 12 | 255 |
| Number of subjects with malalignment (varus or valgus) in both knees | 10 | 179 |
| Number of subjects with malalignment (varus or valgus) in either knee | 11 | 221 |
| Number of subjects with laxity (mild – severe) in either knee | 8 | 109 |
| Average number of days per week in the activity | 4.3 | 1.7 |
| Average number of minutes per week in the activity | 161 | 43.4 |
| % with sitting hours <2, 2 – 4, >4 per day in last 7 days (PASE) | 25, 50, 25 | 19, 55, 24 |
| % who walked < 1, 1 – 2, 2 – 4 hours a day in last 7 days (PASE) | 25, 33, 33 | 40, 40, 11 |

Studies have suggested that in subjects with knee malalignment or laxity, altered tibiofemoral loading could be responsible for biomechanical damage and OA progression[62–64]. A much debated view on the role of the quadriceps in such subjects is that higher muscle strength in malaligned or lax knees *increases* the risk of OA progression[65,66]. If the relationship between lower extremity strength and the risk of OA progression is confounded by knee alignment status, a plausible explanation for the decreasing trend discussed above may be that regular investment in the activity class ([3501–4200), [701–1400)) promotes muscle strength but *advances* OA in subjects with malaligned knees. Though the current guidelines for knee OA management recommend muscle strengthening, our analysis highlights the need for a mechanistic investigation of greater power, given that muscle strength is a modifiable risk factor in OA.

## 4   Discussion

The methods for studying daily activity data obtained from wearable devices have lagged behind the improvements in the accuracy, form factor and manufacturing cost of the devices. An extensively used analysis approach [35,37,67–71] involves binning daily activity minutes into intervals that correspond to activity intensity levels. Results from such analysis depend on study-specific considerations that influence the choice of the histogram bins, as investigated in our earlier work [61]. Moreover, regardless of the specific approach taken, summarization of activity frequency always loses valuable information about the underlying activity-generating processes. For example, walking on an incline at a fixed pace versus walking on level ground at the same pace are distinct in terms of strength and endurance capabilities but their profiles based on intensity bins are similar.

To infer physical function from daily activity trace, it is necessary to derive a representation that conveys information about the "daily activity mix". We define probabilistically distinct segments from daily activity traces as instances of a set of activity classes. We are thus able to transform a sequence of activity counts into a sequence of activity classes, which provide a far more informative view of daily physical activity from the perspective of functional ability. Our approach of unsupervised segmentation and the subsequent definition of notional activity classes allows a function based comparison among subjects without the overhead of obtaining annotated activity traces from subjects. This comparison is similar in spirit to ADLs, is based on objective measurements, and is perhaps the first effort till date to learn and interpret functional capability from wearable device data, within a clinical research use case.

We have illustrated the utility of our daily activity profile in classifying patients into physical function categories and discussed the ramifications of key design choices in a typical machine learning pipeline. Since black-box models have limited use in clinical settings, we focused on techniques that facilitate intuitive interpretations of our classification models. An important observation from our experiments was that even as the activity profile distribution changes over time, the models learned may still be useful for predicting physical function category in the short-to-mid time horizon. Our analysis showed that the change in activity profile from year 4 to year 6 was different in high and low functioning OA subjects, with differential decline in daily average minutes in activity classes that comprised mostly of low to moderate activities. The differential decline between high and low functioning subjects is intuitive and suggests that such classifiers may be useful in remote monitoring applications, at least in the short term and with periodic retraining.

There are two main limitations of the methods described above. First, the mean and SD are likely to be inadequate representations of the activity generating processes as they ignore the temporal relationships between activity counts. Modelling class instances as subsequences generated by a random process has been proposed[72] and may improve detection of activity classes. Second, at the level of the daily activity profile, the time ordering between activity class instances is ignored in the present approach. One way to address both limitations may be to learn the within class and inter-class relationships for a set of daily activity sequences, as a single Bayesian network.

Classifying physical function may be useful in several areas. For example, the evaluation of alternative rehabilitation programs for OA patients recovering from knee surgery requires a comparative assessment of outcomes that include functional recovery. Alternatives to outpatient physical therapy[73] are a topic of active research as care providers focus on improving utilization while assuming financial accountability for the full episode of care. Remote monitoring of physical function in daily living may allow rehabilitation programs to be evaluated in much large populations compared to the sample sizes seen in clinical trials.

Another application may be matching prosthetic devices with the functional requirements and abilities of lower-limb amputees. Currently a match is determined based on a subjective assessment of the disability at baseline, without accounting for real-world activities that a patient engages in. The functional and financial suitability of prostheses (ranging from $45 - $100,00) must be evaluated with greater precision[71], given the prevalence and trend of vascular dysfunction and the projected doubling of limb amputees over the next few decades. With device and application specific tuning, our physical function classifiers may facilitate evidence based matching of disability conditions and specific prostheses.

# 5    Conclusion

An assessment of physical function based on the ability to perform routine tasks in daily living is desirable. The widespread availability of wearable motion sensors in consumer devices offers a way to objectively and unobtrusively record daily activity. We demonstrate a novel approach for deriving a daily activity profile that represents time spent in different tasks encountered in daily living. Classifiers trained on the activity profile are able to correctly anticipate highest and lowest quartile results of clinically used physical performance measures. We recover associations between activity classes and physical performance measures, some of which have support in prior OA research. As shown, capturing physical function as an activity profile can enable remote monitoring of patients' physical function.

# 6  Acknowledgements

We thank Professor Trevor Hastie for his helpful suggestions pertaining to data analysis.

Care Res (Hoboken). 2015;67(2):196–202.

## Supporting Information

**S1 File. Quantitative & Ordinal Response Models**
**S2 File. Effect of changes in the activity profile distribution from year 4 to year 6 on 20MPACE category prediction**